\documentstyle[pre,aps]{revtex}        
\tighten

\begin{document}

\twocolumn[\hsize\textwidth\columnwidth\hsize\csname @twocolumnfalse\endcsname
\title{Connectivity-dependent properties of diluted sytems\\ in a
transfer-matrix description}
\author{
S. L. A. de Queiroz$^a$\footnote{Electronic address:
sldq@if.uff.br} and
 R. B. Stinchcombe$^b$\footnote{Electronic address: stinch@thphys.ox.ac.uk}
 }
\address{
$^a$ Instituto de F\'\i sica, Universidade Federal Fluminense,\\ Avenida
Litor\^anea s/n, Campus da Praia Vermelha, 24210-340 Niter\'oi RJ, Brazil \\
$^b$ Department of Physics, Theoretical Physics, University of Oxford,\\
 1 Keble Road, Oxford OX1 3NP, United Kingdom
}
\date{\today}
\maketitle
\begin{abstract}
We introduce a new approach to connectivity-dependent properties
of diluted systems, which is based on the  transfer-matrix formulation
of the percolation problem. It simultaneously incorporates the
connective properties reflected in non-zero matrix elements and allows
one to use standard random-matrix multiplication techniques. Thus it is
possible to investigate physical processes on the percolation
structure with the high efficiency and precision characteristic of
transfer-matrix methods, while avoiding disconnections. The method
is illustrated for two-dimensional site percolation
by calculating $(i)$ the critical correlation length along 
the strip, and the finite-size longitudinal DC conductivity:
$(ii)$  at the percolation threshold, and $(iii)$ very near
the pure-system limit.  
\end{abstract}

\twocolumn
\narrowtext
\vskip0.5pc]

The transfer-matrix (TM) approach to percolation was pioneered
by Derrida and coworkers\cite{init}. By analysing the possible
combinations of adjacent column states made up of occupied and
unoccupied sites (or bonds), and the allowed connections among 
the latter and to the arbitrary origin of a two-dimensional 
($d=2$) strip, it was possible to
write the TM on the basis of such column states.
The key element in this formulation was the fact that, from the
very structure of the TM, repeated multiplication is tantamount
to the simultaneous generation of all possible connected
configurations that span the strip, each with its proper probabilistic 
weight. Thus the probability of connection to the origin, whose
exponential decay is governed by the correlation length, is 
asymptotically given exactly by the largest eigenvalue of the TM. 
The correlation length
could then be used in a phenomenological
renormalisation calculation\cite{night}, which gave very accurate
results for critical parameters such as the percolation threshold
and correlation-length exponent $\nu$.
It was not clear, however, how one could take advantage of such a direct
and elegant scheme to investigate properties other than the decay of
the probability of connection to the origin.  The obvious
alternative, of building up successive columns by
occupying (or not) each individual site independently, runs into the
problem of disconnections, which is severely aggravated on a strip 
geometry. Up to now, the usual solution has been to generate
configurations site by site, and study quantities which do
not depend on keeping connectivity along the strip, {\it e.g.}   
the moments of the distribution of clusters\cite{sd} for percolation
in $d=2$ and 3. A clever way to get round the effects of
disconnections for random resistor-insulator networks
at  percolation was introduced~\cite{tcond}
by generating individual elements on
long strips (or bars, in $d=3$) with {\it free} edges.
By imposing a fixed voltage drop {\it across} the strip, it was
possible to invoke TM concepts in a step-by-step evaluation of the transverse
conductivity, for which longitudinal disconnections of the resistor
structure are
irrelevant. Finally, in superconductor-resistor networks at the percolation
threshold of superconducting elements, disconnections are in fact
responsible for 
the quantity of interest, which is the residual finite-size 
resistivity; one can then establish periodic boundary conditions
across the strip (in order to minimise finite-width effects)
and estimate the longitudinal resistivity~\cite{suc,suc2}. 
\smallskip\par
Here we introduce a scheme which preserves the connected structure of the
percolating cluster as one sweeps along the strip, and at the same
time relies on standard ideas of random-matrix multiplication. The
latter feature implies that any physical quantity, in addition
to connection probability, can be sampled along the strip through
insertion of its corresponding local realisation. This opens the way
{\it e.g.} to the straightforward treatment of spin-spin correlations in dilute
magnets~\cite{rbs83}, for which so far only approximate TM treatments,
relying on plausible but essentially uncontrollable assumptions,
have been available\cite{dqrbs}. In the new scheme we
enumerate the set of all allowed column combinations, according to
the original TM procedure~\cite{init}, and then build the strip one
full column at a time, by picking a given column's successor at
random but {\it only among those columns that are allowed} by the
connectivity rules ({\it i.e.}, which have a non-zero TM element
linking them to the immediate predecessor). With the
proper assignment of probabilistic weights, as explained below,
this procedure is equivalent to the sampling of connected
configurations implicit in the iteration of the TM.\par
In what follows, we first expose the basic concepts of the method; then
the decay of correlations is calculated and shown to reproduce the 
results given by diagonalisation of the TM. Next we apply the
method to the longitudinal
conductivity of a diluted resistor-insulator network.
For this particular process high-accuracy results exist,
together with some exact ones, which provide a test of the method.
At the percolation point, the new method produces  estimates of the 
conductivity exponent which compare very favourably with those existing 
in the literature. Near the pure-network limit, we obtain the
corrections to the conductivity to first order in defect concentration, in
excellent numerical agreement with analytical results. Finally we point to
possible extensions and generalisations of the present approach.
\smallskip\par
We consider a strip of a square lattice of width $L$ sites, with
periodic boundary conditions across, on which sites may be independently
occupied (unoccupied) with probability $p$ ($1-p$). 
As explained in Ref.~\onlinecite{init}, one builds all possible
column configurations in which at least one site is occupied $and$ linked to
the origin (assumed to be at the far left, say); other sites may be either
occupied and connected, occupied and disconnected, or unoccupied. The TM
element $T_{ij}$ between column configurations
$i$ and $j$ is non-zero only if column $j$
is allowed to succeed column $i$ (meaning: connection to the origin must
be preserved, and illogical situations, such as an occupied and
disconnected site being preceded by a connected one on the same row,
must not occur). One has
\begin{equation}
T_{ij} = p^{N_j} (1-p)^{L-N_j}\ \ \ ,
\label{eq:1}
\end{equation}
\noindent where $N_j$ is the number of occupied sites in $j$.
Our procedure then goes as follows. Assume that the strip has been built
up to a column whose configuration is $i$.
Call $\{j(i)\} \equiv j_1, j_2, \ldots j_{M_i}$ the set of all $j$'s
(a total of $M_i$) allowed to succeed a given $i$.
A segment of length  
\begin{equation}
L_i = \sum_{\{j(i)\}} L_{ij}= \sum_{\{j(i)\}} p^{N_j} (1-p)^{L-N_j}
\label{eq:2}
\end{equation}
\noindent represents the total (conditional) probability to have
a connected configuration succeeding column $i$. Drawing a random
number $ 0 < \epsilon < 1$ from a uniform distribution, the next
column configuration is chosen to be $j_{i_0}$ such that
\begin{equation}
\sum_{j=j_1}^{j_{i_0-1}} L_{ij} < \epsilon L_i <
\sum_{j=j_1}^{j_{i_0}} L_{ij}\ \ \ . 
\label{eq:3}
\end{equation}
\noindent This ensures that the allowed connected configurations come
up with their proper corresponding probabilities. One can then proceed
and generate a column to succeed $j_0$, and so on iteratively.
The only information to be kept in store throughout the process 
is the same as that used in the standard TM formalism:
the location and indices (column occupancy numbers)
of non-zero TM elements. 
\smallskip\par
We now show that for a strip of width $L$,
 the scheme described above gives the same correlation
length, $\xi_L$, as that obtained from diagonalisation of the TM.
One defines $\xi_L(p)$ through the exponential decay of the
probability of connection between columns $0$ and $N$, $P_N(p)$:
\begin{equation}
P_N(p) \sim \exp(-N/\xi_L(p)) \ \ \ .
\label{eq:4}
\end{equation}
\noindent As the process described here is a {\it sequential
sampling} one (as opposed to the parallel updating of mutually 
excluding paths in configuration space, which takes place
in the iteration of the TM), one must consider the appropriate 
quantities to analyse. At each step, when column configuration
$j_{i_0}$
is chosen among $M_i$ to succeed $i$, one is probing one branch of a 
tree structure in the space of column configurations,
and discarding $M_i-1$ others. In order to deal with
this, and produce unbiased samples, the standard procedure is
the weighting of steps
introduced in early simulations of self-avoiding walks~\cite{rr}.
When considering the allowed moves from a point $i$ to the next
(here: in configuration space, as opposed to real space in
Ref.~\onlinecite{rr}) one generates a weight $W_i$ proportional to the
total probability of moving out from that point. In the 
present case, $W_i = L_i$ of Eq.~(\ref{eq:2}). It is easy to see
that $W_i$ is properly normalised, as the denominator is the sum
of probabilities of {\it all} possible succeeeding configurations,
not only connected ones, and that is unity. The total weight of a
given $N-1$- step walk (spanning $N$ points)
is the product $W_1 W_2 \ldots W_N$, to be denoted by ${\cal W}_N$. 
The quantity whose variation with distance is to be studied, in the
present context of sequential sampling, is ${\cal W}_N$; on
universality grounds, it is expected to decay with the same
correlation length as $P_N$ of Eq.~(\ref{eq:4}).\par
For strips of widths $L = 4, \ldots 11$ at $p = 0.592745$,
the best numerical estimate of the appropriate percolation threshold 
to our knowledge~\cite{ziff}, we have generated large
numbers ($N_{samp}$) of independent connected configurations between
the origin and column $N_0$. The weights of the $N_{samp}$ configurations
up to respectively  columns $N_0 - \Delta$ and $N_0$ were summed
to produce the estimates ${\overline{\cal W}_{N_0 - \Delta}} = 
(\sum_{k=1}^{N_{samp}}{\cal W}_{N_0 - \Delta}^{(k)}) / N_{samp}$ and
${\overline{\cal W}_{N_0}} = 
(\sum_{k=1}^{N_{samp}}{\cal W}_{N_0}^{(k)}) / N_{samp}$ . An estimate
of the correlation length is then given by
\begin{equation}
{\Delta \over \xi} = - \ln \left({\overline{{\cal W}}_{N_0}
 \over \overline{{\cal W}}_{N_0 -\Delta}}\right)\ \ \ . 
\label{eq:5}
\end{equation}
\noindent Finally, we have repeated the process $n_s$ times with distinct
random-number sequences, in order to estimate fluctuations. We used
$N_{samp} = 10^5$, $N_0 = 20$, $\Delta = 10$, $n_s = 100$.  Our results
are displayed in Table 1,
in the form of estimates for the critical decay-of-correlations 
exponent $\eta$, through the identity $\eta = L/\pi\xi(p_c)$ 
given by conformal invariance~\cite{cardy}. These are
 to be compared with those, also in Table 1, obtained from the
largest eigenvalue of the TM. The values of $N_0$ and $\Delta$ were chosen
 bearing in mind that, both from general finite-size scaling ideas and 
from previous results for percolation~\cite{init}, it is known that 
the correlation length at criticality must be of order $L$. 
 The convergence of finite-width results
towards the value given by conformal invariance~\cite{cardy},
$\eta = 5/24 = 0.208333\ldots$ has been investigated elsewhere~\cite{dq95}.
For our present purposes the relevant comparison is between the
columns of Table 1, which shows the soundness of the proposed scheme.
As expected from the theory of normal distributions, fluctuations
shrink with $\left(N_{samp}\right)^{-1/2}$, because it is there that the
accumulation of sample weights leading to self-averaging takes place,
but remain approximately constant with
$n_s$. The amount of computational time involved is linearly 
proportional to $n_s \times N_{samp}$, thus one can produce more 
accurate results by increasing $N_{samp}$ while reducing $n_s$ to one;
in this limit, the width of the error bar for the single 
(presumably very precise) central estimate
can be extrapolated from those obtained for large $n_s$ and 
correspondingly smaller $N_{samp}$, via the $\left(N_{samp}\right)^{-1/2}$
dependence. Since our goal here is to demonstrate the feasibility
of the proposed approach, rather than refining numerical values, we
did not pursue this line systematically. 
\smallskip\par  
We now show results for finite-size conductivity $\sigma_L (p_c)$
at the percolation threshold of a resistor-insulator network.
 From finite-size scaling, this is expected to vary with strip width as
\begin{equation}
\sigma_L (p_c) \sim L^{-t/\nu} \ \ \ ,
\label{eq:6}
\end{equation}
\noindent where the best available estimate for the exponent is
$t/\nu = 0.9745 \pm 0.0015$~\cite{suc2}. We have generated samples of
site-diluted resistor networks (where a bond is a resistor if it connects
two occupied sites, and an insulator otherwise), according to the 
procedure delineated above. Now, since the quantity to sample
(average conductivity per bond) is
naturally accumulated as one proceeds along the strip (instead of
decaying exponentially, as is the case with connection probabilities),
one does not have to be concerned with weights: it suffices
to generate very long samples, and each column configuration will come
up with its good weight.\par
Conductivities have been calculated by Fogelholm's node deletion 
algorithm~\cite{fogel}. This is very efficient on a strip geometry since
it depends on keeping track only of at most $L(L-1)/2$ links among sites, plus
$2L$ links to the origin.\par 
Table 2 shows our data for $L = 3$--11 where for each strip width 100 
independent samples, each of length $10^5$ columns, were generated. Error
bars reflect deviations among different samples. A least-squares fit to a
log-log plot of the data in Table 2 gives $t/\nu = 1.005 \pm 0.002$, with
an accumulated $\chi^2$ per degree of freedom (DOF) $= 1.0$, an estimate
which is 3\% above the accepted value~\cite{suc2}, with apparently 
non-overlapping error bars. Before accepting this at face value, 
some remarks are in order.\par
Firstly, we perform a similar fit to the resistivity ($\rho_L$) data on the 
site superconductivity problem, shown in Table I of Ref.~\onlinecite{suc2}. 
In $d=2$ the superconducting exponent $s/\nu$
in $\rho_L \sim L^{-s/\nu}$ is the same as $t/\nu$, by duality
(see Refs.~\onlinecite{suc,suc2}). Also, the same (periodic) boundary 
conditions across the strip were used there, 
as opposed to free ones for previous conductivity  studies~\cite{tcond}.
This is important when one wishes to compare purely finite-size effects
between two sets of data. Those authors
simulated strips $10^4$ times as long as we did, thus the fact that their
error bars are two orders of magnitude smaller than our own indicates that
both methods have the same intrinsic accuracy. We then turn to comparison of
systematic errors. 
For the {\it same} range $3 \leq L \leq 11$, the data
of Ref.~\onlinecite{suc2}
 give $t/\nu \simeq 0.913$, though with a very large
$\chi^2$/DOF  which partly reflects the greater 
accuracy of individual data in Ref.~\onlinecite{suc2}, as well as the need
to take corrections to scaling into account. Assuming a power-law correction
with exponent $\omega$, always for $3 \leq L \leq 11$,
fits of $\rho_L L^{t/\nu}$ to $a +b/L^{\omega}$
for $0.94 \leq t/\nu \leq 1.02$
show that $\chi^2$/DOF indeed has a minimum value at
$t/\nu \sim 0.982$--0.987 when $\omega$
is kept constant at 1.2 -- 1.4 (see Ref.~\onlinecite{suc2}). The amplitude
$b$ varies monotonically between $-0.1$ and $-0.6$. A similar analysis
of our own data shows that $\chi^2$/DOF has a gentle maximum at
$t/\nu \sim 0.95$ and a minimum at  $t/\nu \sim 1.025$. The
amplitude $b$ starts from $0.44$ at $t/\nu = 0.94$ and decreases 
monotonically, crossing zero at $t/\nu \sim 1.005$. Varying $\omega$ 
along a wider interval, between 1 and 2.5, does not produce any significant 
change.
\par
Thus, for similar strip widths and comparable amounts of computational effort,
our method generates data of comparable quality to other 
authors'. It seems that for critical conductivity studies in two dimensions
one has to reach very
large widths, of order 40 sites~\cite{tcond,suc,suc2}, 
before asymptotic behaviour sets in. While
this could be  done in Refs.~\onlinecite{tcond,suc,suc2}, the nature of the
present algorithm is such that the exponential growth, with strip width,
of the number of configurations to be stored is the main obstacle to going 
further than $L=11$. However, from past experience~\cite{dqrbs} we expect 
such upper limit not to be as stringent for {\it e.g.} diluted magnets.
\smallskip\par
We have also studied resistor networks for very low impurity (insulator)
concentrations ($1-p$), where it has been predicted~\cite{wl} 
that conductivity must vary as
\begin{equation}
\sigma(p)/\sigma(1) = 1 - \pi (1-p) + \pi (1-p)^2/2\ \ \ .
\label{eq:7}
\end{equation}
\noindent By using finite-size  considerations pertaining to low concentrations
of impurities in a 
cylindrical geometry, we have shown that the finite-size conductivity is
\begin{equation}
\sigma_L(p) = \sigma_{\infty}(p) + a(p)/L^2 + O(1/L^4)\ \ \ ,
\label{eq:8}
\end{equation}
\noindent  where $\sigma_{\infty}(p)$ is given by Eq.~(\ref{eq:7}) and
$a(p) = -(1-p)\pi^3/6$. Our data for the normalised conductivity at
$p =0.999$ (where Eq.~(\ref{eq:8}) gives $\sigma_{\infty}(p) = 0.99686 \ldots$
and $a(p)= -5.17\times 10^{-3}$ ) and $3 \leq L \leq 11$ are shown in Table 3,
where for each strip width 100 
independent samples, each of length $10^5$ columns, were generated. Error
bars reflect deviations among different samples.  
A least-squares fit of our finite-size data gives $\sigma_{\infty} =
0.99685 \pm 0.00005$, where the small error bar reflects the fit's overall
smoothness, and $a=(-6.0 \pm 1.7)\times 10^{-3}$, in very good agreement with
the theoretical prediction.
\smallskip\par
We have proposed and illustrated a straightforward scheme for diluted systems, 
in which
a transfer-matrix approach can be implemented without giving rise to
longitudinal disconnections along a strip. Previous treatments either were
restricted to the calculation of the decay of connection
probability~\cite{init}, or could be carried out independent of disconnections,
owing to particular geometric~\cite{tcond} or physical~\cite{suc,suc2} features,
or else were forced to rely on essentially uncontrollable assumptions on the
commutation of TMs associated to distinct dilution configurations~\cite{dqrbs}.
Extensions of the present work to dilute magnets~\cite{rbs83,dqrbs} are now
being considered. Further applications would be to the anomalous thermal 
behaviour of Fe(110) submonolayers on W(110)~\cite{grad},
and to frustrated percolation~\cite{coniglio},
a  problem related to glass-formation processes.
\medskip\par
SLAdQ thanks the Department of Theoretical Physics
at Oxford, where most of this work was carried out, for the hospitality, and
the cooperation agreement between Conselho Nacional de Desenvolvimento
Cient\'\i fico e Tecnol\'ogico (CNPq) and  
the Royal Society for funding his visit. Research of SLAdQ
is partially supported by the Brazilian agencies Minist\'erio da Ci\^encia
e Tecnologia, Conselho Nacional
de Desenvolvimento Cient\'\i fico e Tecnol\'ogico and Coordena\c c\~ao de
Aperfei\c coamento de Pessoal de Ensino Superior.

\vskip 0.6cm
\begin{table}
\caption{Estimates of $\eta = L/\pi \xi(p_c)$ at $p=0.592745$;\\ averages
over $n_s = 100$ distinct sequences, each of $N_{samp} = 10^5$ accumulated
weights
.}
\vskip 0.4cm 
 \halign to \hsize{\hskip 0.5cm
\hfil#\quad\hfil&\quad\hfil#\quad\hfil&&\hfil#\quad\hfil\cr
    $L$ & $\eta$ (this work) &  $\eta$ (TM) \cr \noalign{\smallskip} 

4   &   $0.21255 \pm 0.00020$  &    0.2125576128 \cr   
5   &   $0.21142 \pm 0.00026$  &    0.2114673276 \cr
6   &   $0.21069 \pm 0.00029$  &    0.2107370714 \cr
7   &   $0.21016 \pm 0.00039$  &    0.2102232886 \cr
8   &   $0.20979 \pm 0.00043$  &    0.2098564767 \cr
9   &   $0.20954 \pm 0.00044$  &    0.2095868033 \cr
10  &   $0.20930 \pm 0.00047$  &    0.2093833099 \cr
11  &   $0.20915 \pm 0.00055$  &    0.2092261631 \cr
}
\end{table}

\begin{table}
\caption{Estimates of $\sigma_L(p_c)$ at $p=0.592745$; averages
over $n_s = 100$ distinct strips, each of length $10^5$  columns
.}
\vskip 0.4cm 
 \halign to \hsize{\hskip 1.5cm
\hfil#\quad\hfil&\quad\hfil#\quad\hfil\cr
    $L$ & $\sigma_L$  \cr \noalign{\smallskip} 

  3   &  $0.34435  \pm   0.00063$ \cr  
  4   &  $0.25931  \pm   0.00054$ \cr     
  5   &  $0.20718  \pm   0.00053$ \cr
  6   &  $0.17241  \pm   0.00047$ \cr
  7   &  $0.14749  \pm   0.00045$ \cr
  8   &  $0.12881  \pm   0.00038$ \cr         
  9   &  $0.11431  \pm   0.00042$ \cr
 10   &  $0.10271  \pm   0.00035$ \cr
 11   &  $0.09323  \pm   0.00033$ \cr
}
\end{table} 
\begin{table}
\caption{Estimates of $\sigma_L(p)$ at $p=0.999$; averages\\
over $n_s = 100$ distinct strips, each of length $10^5$ columns.
 Extr.: extrapolation against $1/L^2$ (see text).
Expected: Eq.~(\protect{\ref{eq:7}}). 
}
\vskip 0.4cm 
 \halign to \hsize{\hskip 1.5cm
\hfil#\quad\hfil&\quad\hfil#\quad\hfil\cr
    $L$ & $\sigma_L$  \cr \noalign{\smallskip} 
 
 3   &   $0.99618  \pm    0.00019$ \cr
 4   &   $0.99648  \pm    0.00016$ \cr
 5   &   $0.99661  \pm    0.00013$ \cr
 6   &   $0.99669  \pm    0.00012$ \cr
 7   &   $0.99673  \pm    0.00011$ \cr
 8   &   $0.996753  \pm   0.000099$ \cr
 9   &   $0.996775  \pm   0.000099$ \cr
10   &   $0.996791  \pm   0.000095$ \cr
11   &   $0.996804    \pm 0.000092$ \cr
Extr. &  $0.99685   \pm 0.00005$ \cr
Expected &  $0.99686\ldots$ \cr
}
\end{table}

\end{document}